\documentclass{article}

\usepackage{multicol} 
\usepackage{latexsym}
\usepackage[svgnames]{xcolor} 

\usepackage{times} 

\usepackage{graphicx} 
\graphicspath{{figures/}} 
\usepackage{booktabs} 
\usepackage[font=small,labelfont=bf]{caption} 
\usepackage{amsfonts, amsmath, amsthm, amssymb} 
\usepackage{wrapfig} 

\title{Special functions with mod $n$ symmetry and kaleidoscope of quantum coherent states}

\author{Ayg\"{u}l Ko\c{c}ak$^1$ and  Oktay K. Pashaev$^2$\\
Department of Mathematics , Izmir Institute of Technology, \\Izmir, 35430, Turkey\\
$^1$aygulkocak@iyte.edu.tr, $^2$oktaypashaev@iyte.edu.tr}

\begin{document}

\maketitle

\begin{abstract}
The set of mod $n$ functions associated with primitive roots of unity and discrete Fourier transform is introduced. These functions naturally appear in description of superposition of coherent states related with regular polygon, which we call kaleidoscope of quantum coherent states. Displacement operators for kaleidoscope states are obtained by mod $n$ exponential functions with operator argument and non-commutative addition formulas. Normalization constants, average number of photons, Heinsenberg uncertainty relations and coordinate representation of wave functions with mod $n$ symmetry are expressed in a compact form by these functions.
\end{abstract}

\section{Introduction}

In description of superposition of Glauber coherent states associated with regular n polygon and roots of unity$\cite{kocakpashaev}$,  the main characteristics of the states as normalization factors, coordinate representations, uncertainty relations, etc. appear as  infinite sums of exponential functions with mod $n$ symmetry $\cite{bb},\cite{Stoler},\cite{Spiridonov}$. For $n$=2 the  exponential functions with mod $2$ symmetry are just the  hyperbolic functions, this is why, generalizations to arbitrary  $n$  were  called as generalized hyperbolic functions$\cite{ungar}$.
Here, for description of kaleidoscope of coherent states we introduce generic special functions with mod $n$ symmetry and call them shortly as the mod $n$ functions, so that the generalized hyperbolic functions become a particular case.
As application of these functions, we derive displacement operators for kaleidoscope states by mod $n$ exponential functions with operator argument and non-commutative addition formulas. Normalization factors, number of photons and uncertainty relation we express by mod $n$ exponential functions. Then we calculate
the generating function for kaleidoscope states by mod $n$ Gaussian exponential function and $n$-paricle states in coordinate representation by mod $n$ Hermite polynomials.

The cat states as orthogonal coherent states, representing simplest kaleidoscope superposition  with mod $2$ symmetry, have been used for description of qubit unit of quantum information in quantum optics. But for quantum information processing the ternary and the quaternary systems, with base 3 and 4  for position notation could be more efficient. This requires generalization of cat states to trinity states and  quartet states with mod $3$ and mod $4$ symmetry,  providing units of quantum information as qutrit and ququad.  More generally, mod $n$ kaleidoscope of states with arbitrary $n$, furnishes orthonormal  basis for description of generic qudit unit of quantum information. This is why for such quantum states, physical and information characteristics as entanglement, entropy and randomness, would be naturally represented by mod $n$ functions.

\subsection{Scale and Phase Invariance}
The set of mod n functions satisfy self-similarity property under rotations. A function is said to be scale-invariant if it satisfies following property;
\begin{eqnarray} \nonumber
f(\lambda z)=\lambda^{d}f(z) \, ,
\end{eqnarray}
for some choice of exponent $d\in\mathbb{R}$ and fixed scale factor $\lambda>0$, which can be taken as a length or size of re-scaling. If $\lambda=e^{i\varphi}$ and as follows $|\lambda|=1$, then this formula gives
\begin{eqnarray} \nonumber
f(e^{i\varphi} z)=e^{i\varphi d}f(z).
\end{eqnarray}
In this case, rotation of argument $z$ to angle $\varphi$ implies rotation of function $f$ to angle $\varphi d,$ and scale invariance becomes rotational or phase(gauge) invariance. If $\lambda=q^2$ is the primitive root of unity
\begin{eqnarray} \nonumber
q^{2n}=1
\end{eqnarray}
so that $q^2=e^{i\frac{2\pi}{n}}$, then
\begin{eqnarray} \nonumber
f(e^{i\frac{2\pi}{n}} z)=e^{i\frac{2\pi}{n}d}f(z) \, .
\end{eqnarray}
This means that rotation of argument $z$ to angle $\frac{2\pi}{n}$ of $n-$sided polygon, leads to rotation of $f$ on $d-$times of this angle. We call this as discrete phase gauge invariant function, with order $d$. Simplest example of phase invariant functions is given by even and odd functions with $q^4=1$;
\begin{eqnarray} \nonumber
f_{even}(q^2x)=f_{even}(x) \,\, , \,\, f_{odd}(q^2x)=q^2f_{odd}(x)
\end{eqnarray}
where $\lambda=q^2=-1$ and $d=0\,,\,d=1$ respectively.
\section{Mod n functions}
For calculation of normalization constants and average number of photons in kaleidoscope of quantum coherent states, we introduce mod n functions. For $q^{2n}=1$ primitive root of unity, we consider n values of argument $\displaystyle{x,q^{2}x,q^{4}x,...,q^{2(n-1)}x}$ as rotated by angle $\frac{2\pi}{n}$ and associated with vertices of regular polygon. Mod $n$ functions are defined by relation
\begin{eqnarray} \label{disfourier}
\left[
  \begin{array}{c}
   f_{0}(x) \\
   f_{1}(x)  \\
   f_{2}(x)  \\
   \vdots \\
    f_{n-1}(x) \\
  \end{array}
\right]=\frac{1}{n}\left[
                    \begin{array}{cccccccc}
1 & 1 & 1 & ... & 1 \\
1 & \overline{q}\,^2 & \overline{q}\,^4 & ... & \overline{q}\,^{2(n-1)} \\
1 & \overline{q}\,^4 & \overline{q}\,^8 & ... & \overline{q}\,^{4(n-1)}\\
\vdots & \vdots & \vdots & \ddots & \vdots \\
1 &\overline{q}\,^{2(n-1)}  & \overline{q}\,^{4(n-1)} & ...  & \overline{q}\,^{2(n-1)^2} \\
                             \end{array}
                           \right]\left[
  \begin{array}{c}
    f(x) \\
    f(q^2x) \\
    f(q^4x)  \\
    \vdots \\
    f(q^{2(n-1)}x) \\
  \end{array}
\right] \, ,
\end{eqnarray}
where transformation matrix is discrete Fourier transformation, related with unitary generalized Hadamard gate matrix $\hat{H}.$ Every mod n function is a superposition of functions with these arguments, such that addition of coefficients for  $\displaystyle{f(q^{2}x),f(q^{4}x),...,f(q^{2(n-1)}x)}$
are equal to zero, due to
\begin{eqnarray}
 1+\overline{q}\,^{2k}+\overline{q}\,^{4k}+...+\overline{q}\,^{2(n-1)k}=0 \, , 1\leq k \leq n-1 \, . \nonumber
\end{eqnarray}
By inverting transformation $(\ref{disfourier})$, it is evident that arbitrary function $f(x)$ can be written as a superposition of mod $n$ functions
\begin{eqnarray} \label{superpositionofnfunction}
f_{k}(x)=\frac{1}{n}\sum_{s=0}^{n-1} \overline{q}^{2sk}f({q}^{2s}x)
\end{eqnarray}
in a unique way, $\displaystyle{f(x)=\sum_{k=0}^{n-1}f_{k}(x) \,\, .}$

\subsection{Mod n exponential functions}
Most important example of this expansion is given by exponential function,
\begin{eqnarray} \nonumber
&e^{x}&=\sum _{m=0}^{\infty} \frac{x^m}{m!}=\sum _{k=0}^{\infty} \frac{x^{nk}}{nk!}+
\sum _{k=0}^{\infty} \frac{x^{nk+1}}{(nk+1)!}  +
\sum _{k=0}^{\infty} \frac{x^{nk+2}}{(nk+2)!}  + ... +
\sum _{k=0}^{\infty} \frac{x^{nk+(n-1)}}{(nk+(n-1))!} .
\end{eqnarray}
Every sum here represents mod n exponential function $f_{s}(x)\equiv\, _{s}e^{x}(mod\,n)$,$0\leq s \leq n-1$,
\begin{eqnarray}
 _{s}e^{x}(mod\,n)\equiv\sum_{k=0}^{\infty}\frac{x^{nk+s}}{(nk+s)!}. \nonumber
\end{eqnarray}
Due to $(\ref{superpositionofnfunction})$ for $f(x)=e^x,$ they can be expressed as superposition of standard exponentials by discrete Fourier transformation,
\begin{eqnarray}
_{s}e^{x}(mod\,n)=\frac{1}{n}\sum_{k=0}^{n-1} \overline{q}^{2sk}e^{{q}^{2k}x} \,\,. \nonumber
\end{eqnarray}
For n=2, the mod 2 exponential functions coincide with hyperbolic functions:
\begin{eqnarray} \nonumber
_{0}e^{x}=\cosh x \quad , \quad  _{1}e^{x}=\sinh x.
\end{eqnarray}
This is why, it is natural to call mod n exponential functions for arbitrary n, as generalized hyperbolic functions$\cite{ungar}$.
\\The derivative operator is acting on mod n exponential functions in following way,
\begin{eqnarray*}
\frac{d}{dx}\,_{0}e^{x}=\,_{n-1}e^{x} \quad , \quad\frac{d}{dx}\,_{s}e^{x}=\,_{s-1}e^{x} \quad, 1\leq s \leq n-1 .
\end{eqnarray*}
Applying this derivative n times, we find that function $f_s(x)=\, _{s}e^{x}$ is a solution of ordinary differential equation of degree $n$,
\begin{equation} \nonumber
 f^{(n)}_{s}= f_{s} \, ,\quad\textrm{where} \,\,  0\leq s \leq n-1 , \label{difeqn}
\end{equation}
with initial values: $f^{(s)}_s(0)=1, f_s(0)=f'_s(0)=...=f^{(s-1)}_s (0)=f^{(s+1)}_s (0)=...= f^{(n-1)}_s (0)=0.$
This differential equation is the eigenvalue problem $\hat{a}^n f = f $ for annihilation operator $\hat{a}=\frac{d}{dz}$ in the Fock-Bargmann representation, acting on analytic function $f=f(z)$.

\section{Displacement operators for kaleidoscope states}
As a first application of mod $n$ exponential functions, we consider displacement operators for kaleidoscope of coherent states. Application of this displacement operator to vacuum state requires factorization of mod n exponential functions with operator argument. Below we describe in details this factorization for mod 2 exponential functions.

\subsection{Factorization of mod $2$ exponential functions with operator argument}
As well known exponential function with operator argument can be factorized in the form
\begin{eqnarray} \label{exponentialfactorization}
e^{\hat{A}+\hat{B}}=e^{\hat{A}}e^{\hat{B}}e^{-\frac{1}{2}[\hat{A},\hat{B}]} \, ,
\end{eqnarray}
where $\hat{A}$ and $\hat{B}$ are c-commutative: $[\hat{A},[\hat{A},\hat{B}]]=[\hat{B},[\hat{A},\hat{B}]]=0.$

Here, we derive factorization formulas for mod $2$ exponential functions:
Let $\hat{A}$ and $\hat{B}$ are two c-commutative operators, then
\begin{eqnarray}\label{backerhousdorfftwoexponential1}
_0 e^{\hat{A}+\hat{B}}&=&\left(\,_0 e^{\hat{A}}\,_0 e^{\hat{B}}+\,_1 e^{\hat{A}}\,_1 e^{\hat{B}}\right)e^{-\frac{1}{2}[\hat{A},\hat{B}]} \, , \\
 \label{backerhousdorfftwoexponential2}
_1 e^{\hat{A}+\hat{B}}&=&\left(\,_0 e^{\hat{A}}\,_1 e^{\hat{B}}+\,_1 e^{\hat{A}}\,_0 e^{\hat{B}}\right)e^{-\frac{1}{2}[\hat{A},\hat{B}]}.
\end{eqnarray}
Factorization formula $(\ref{exponentialfactorization})$ gives $q-$commutative relation between operators $e^{\hat{A}}$ and $e^{\hat{B}},$
\begin{eqnarray} \nonumber
e^{\hat{A}}e^{\hat{B}}=e^{[\hat{A},\hat{B}]}e^{\hat{B}}e^{\hat{A}}= q e^{\hat{B}}e^{\hat{A}} \,.
\end{eqnarray}
We have analogue of this formula for mod $2$ exponential functions.
For operators $\hat{A}$ and  $\hat{B}$ such that $[\hat{A},[\hat{A},\hat{B}]]=[\hat{B},[\hat{A},\hat{B}]]=0$ following identities hold
\begin{eqnarray*}
\,_0 e^{\hat{A}}\,_0 e^{\hat{B}}&=&\,_0 e^{\hat{B}}\,_0 e^{\hat{A}}\,_0e^{[\hat{A},\hat{B}]}
+\,_1 e^{\hat{B}}\,_1 e^{\hat{A}}\,_1 e^{[\hat{A},\hat{B}]} ,\\
\,_1 e^{\hat{A}}\,_1 e^{\hat{B}}&=&\,_1 e^{\hat{B}}\,_1 e^{\hat{A}}\,_0e^{[\hat{A},\hat{B}]}
+\,_0 e^{\hat{B}}\,_0 e^{\hat{A}}\,_1e^{[\hat{A},\hat{B}]}, \\
\,_0 e^{\hat{A}}\,_1 e^{\hat{B}}&=&\,_1 e^{\hat{B}}\,_0 e^{\hat{A}}\,_0e^{[\hat{A},\hat{B}]}
+\,_0 e^{\hat{B}}\,_1 e^{\hat{A}}\,_1e^{[\hat{A},\hat{B}]}, \\
\,_1 e^{\hat{A}}\,_0 e^{\hat{B}}&=&\,_0 e^{\hat{B}}\,_1 e^{\hat{A}}\,_0e^{[\hat{A},\hat{B}]}
+\,_1 e^{\hat{B}}\,_0 e^{\hat{A}}\,_1e^{[\hat{A},\hat{B}]}.
\end{eqnarray*}
These identities can be rewritten in terms of hyperbolic functions of operator argument:
\begin{eqnarray*}
\cosh \hat{A}\cosh \hat{B}&=& \cosh\hat{B} \cosh \hat{A} \cosh[\hat{A},\hat{B}]+ \sinh \hat{B}  \sinh \hat{A} \sinh[\hat{A},\hat{B}], \\
\sinh \hat{A}\sinh \hat{B}&=&\sinh \hat{B} \sinh \hat{A} \cosh[\hat{A},\hat{B}]+ \cosh \hat{B}  \cosh \hat{A} \sinh[\hat{A},\hat{B}], \\
\cosh \hat{A}\sinh \hat{B}&=& \sinh \hat{B} \cosh \hat{A} \cosh[\hat{A},\hat{B}]+ \cosh \hat{B}  \sinh \hat{A} \sinh[\hat{A},\hat{B}], \\
\sinh \hat{A}\cosh \hat{B}&=& \cosh \hat{B} \sinh \hat{A} \cosh[\hat{A},\hat{B}]+ \sinh \hat{B}  \cosh \hat{A} \sinh[\hat{A},\hat{B}].
\end{eqnarray*}
Formulas $(\ref{backerhousdorfftwoexponential1})$ and $(\ref{backerhousdorfftwoexponential2})$, imply also addition formulas for hyperbolic functions of operator argument:
\begin{eqnarray} \label{hyperbolicoperatorargumentaddition1}
\cosh \left(\hat{A}+\hat{B}\right)&=& \left(\cosh \hat{A}\cosh\hat{B}+ \sinh \hat{A}\sinh \hat{B} \right) e^{-\frac{1}{2}[\hat{A},\hat{B}]} ,  \\ \label{hyperbolicoperatorargumentaddition2}
\cosh \left(\hat{A}-\hat{B}\right)&=&\left(\cosh \hat{A}\cosh\hat{B}- \sinh \hat{A}\sinh \hat{B} \right) e^{\frac{1}{2}[\hat{A},\hat{B}]},\\ \label{hyperbolicoperatorargumentaddition3}
\sinh \left(\hat{A}+\hat{B}\right)&=& \left(\sinh \hat{A}\cosh\hat{B}+\sinh \hat{B}\cosh\hat{A}  \right) e^{-\frac{1}{2}[\hat{A},\hat{B}]},\\ \label{hyperbolicoperatorargumentaddition4}
\sinh \left(\hat{A}-\hat{B}\right)&=& \left(\sinh \hat{A}\cosh\hat{B}-\sinh \hat{B}\cosh\hat{A}   \right) e^{\frac{1}{2}[\hat{A},\hat{B}]} .
\end{eqnarray}
For special case, when $[\hat{A},\hat{B}]=0$, these addition formulas reduce to usual formulas for hyperbolic functions. In the next section, we apply these formulas for factorization of displacement operators for Schr\"{o}dinger's cat states.
\subsection{Mod n displacement operator}
\subsubsection{Mod 2 case}
The displacement operators $D(\mp\alpha)$ as exponential function of operator argument
\begin{eqnarray} \nonumber
D(\mp\alpha)= e^{ \mp \alpha \hat{a}^{\dagger} \pm \bar{\alpha} \hat{a}}= e^{ -\frac{1}{2}|\alpha|^2 } e^{ \mp\alpha \hat{a}^{\dagger} }  e^{\pm \bar{\alpha}\hat{a}} \, ,
\end{eqnarray}
generate coherent states $|\mp \alpha \rangle$;
\begin{eqnarray}
| \mp \alpha \rangle = D( \mp \alpha) | 0 \rangle \, . \nonumber
\end{eqnarray}
Superpositions of these states as the cat states can be created by mod $2$ displacement operators $_{0}D(\alpha)$ and $_{1}D(\alpha):$
\begin{eqnarray}  \nonumber
|\tilde{0}\rangle_{\alpha}&=& \frac{|\alpha\rangle+|-\alpha\rangle}{2}= \left(\frac{D(\alpha)+D(-\alpha)}{2}\right)|0\rangle=\,_{0}D(\alpha)|0\rangle \, ,\\ \nonumber
|\tilde{1}\rangle_{\alpha}&=& \frac{|\alpha\rangle-|-\alpha\rangle}{2}=\left(\frac{D(\alpha)-D(-\alpha)}{2}\right) |0\rangle=\,_{1}D(\alpha)|0\rangle \, .
\end{eqnarray}
Due to identities $(\ref{hyperbolicoperatorargumentaddition1})-(\ref{hyperbolicoperatorargumentaddition4}),$ these operators can be written as
\begin{eqnarray} \nonumber
_{0}D(\alpha)&=&e^{ -\frac{1}{2}|\alpha|^2 }(\cosh\alpha \hat{a}^{\dagger}\cosh \alpha \hat{a}-\sinh\alpha \hat{a}^{\dagger} \sinh\alpha \hat{a} ) \, ,\\ \nonumber
_{1}D(\alpha)&=&e^{ -\frac{1}{2}|\alpha|^2 }(\sinh\alpha \hat{a}^{\dagger}\cosh \alpha \hat{a}+\cosh\alpha \hat{a}^{\dagger} \sinh\alpha \hat{a})  \, ,
\end{eqnarray}
and the cat states become
\begin{eqnarray} \label{notnormalizedcat1}
|\tilde{0}\rangle_{\alpha}&=&_{0}D(\alpha)|0\rangle=e^{ -\frac{1}{2}|\alpha|^2 }\cosh\alpha \hat{a}^{\dagger}|0\rangle\, ,\\  \label{notnormalizedcat2}
|\tilde{1}\rangle_{\alpha}&=&_{1}D(\alpha)|0\rangle=e^{ -\frac{1}{2}|\alpha|^2 }\sinh\alpha \hat{a}^{\dagger}|0\rangle    \, .
\end{eqnarray}
Normalization of these states is given in $(\ref{cat1}),(\ref{cat2}).$
\subsubsection{Mod 3 case}
Displacement operators defined by mod $3$ operator valued exponential functions, determine the kaleidoscope of three states(the trinity states) as,
\begin{eqnarray*}
_{0}D(\alpha)=\frac{D(\alpha)+D(q^{2}\alpha)+D(q^{4}\alpha)}{3} &\Rightarrow& |\tilde {0}\rangle_{\alpha}= \,_{0}D(\alpha)|0\rangle  \, ,\\
_{1}D(\alpha)=\frac{D(\alpha)+ \overline{q}^{2}D(q^{2}\alpha)+ \overline{q}^{4}D(q^{4}\alpha)}{3}&\Rightarrow& |\tilde {1}\rangle_{\alpha}=\,_{1}D(\alpha)|0\rangle \, ,\\
_{2}D(\alpha)=\frac{D(\alpha)+ \overline{q}^{4}D(q^{2}\alpha)+ \overline{q}^{2}D(q^{4}\alpha)}{3}&\Rightarrow& |\tilde {2}\rangle_{\alpha}=\,_{2}D(\alpha)|0\rangle \, .
\end{eqnarray*}
\subsubsection{Mod n kaleidoscope states}
The above construction can be generalized to arbitrary mod $n$ case, $q^{2n}=1,$ described by displacement operators
\begin{eqnarray} \nonumber
_{k}D(\alpha)=\frac{1}{n}\sum _{j=0}^{n-1}\overline{q}^{2jk}D(q^{2j}\alpha)\,\, , 0\leq k \leq n-1\, .
\end{eqnarray}
Acting to vacuum state, they produce kaleidoscope of coherent states
\begin{eqnarray} \nonumber
|\tilde {k}\rangle_{{\alpha}}=\,_{k}D(\alpha)|0\rangle \,.
\end{eqnarray}
Operators $\,_{k}D(\alpha)$ are not unitary, this is why, the states $|\tilde {k}\rangle_{{\alpha}}$ are not normalized. In following sections, we show that normalization of these states can be written in a compact form by using mod $n$ exponential functions with argument $|\alpha|^2.$
\section{Generating function for mod $n$ Hermite polynomials}
\subsection{Mod 2 Hermite polynomials}
Coordinate representation of cat states $(\ref{coordinatecat1}),(\ref{coordinateca2})$ is related with generating functions for Hermite polynomials of even $H_{2k}(x)$ and odd $H_{2k+1}(x)$ order as;
\begin{eqnarray} \nonumber
&&\sum _{k=0}^{\infty}\frac{z^{2k}}{(2k)!}H_{2k}(x)=\,e^{-z^2}\cosh(2zx)=\, _{0}e^{-z^2+2zx} \, , \\ \nonumber
&&\sum _{k=0}^{\infty}\frac{z^{2k+1}}{(2k+1)!}H_{2k+1}(x)=\,e^{-z^2}\sinh(2zx)=\, _{1}e^{-z^2+2zx} \, .
\end{eqnarray}
\subsection{Mod 3 Hermite polynomials}
In a similar way, coordinate representation of trinity states is related with mod 3 exponential functions, which are generating functions for $H_{3k}(x),H_{3k+1}(x)$ and  $H_{3k+2}(x)$ Hermite polynomials;
\begin{eqnarray*}
&&\sum _{k=0}^{\infty}\frac{z^{3k}}{(3k)!}H_{3k}(x)=\, _{0}e^{-z^2+2zx}=\frac{1}{3}\left(  e^{-z^2+2zx}+2e^{\frac{z^2}{2}-zx}\cos \left(\frac{\sqrt{3}}{2}(z^2+2zx)\right) \right) \, , \\ \nonumber
&&\sum _{k=0}^{\infty}\frac{z^{3k+1}}{(3k+1)!}H_{3k+1}(x)=\, _{1}e^{-z^2+2zx}=\frac{1}{3}\left(  e^{-z^2+2zx}+2e^{\frac{z^2}{2}-zx}\cos \left(\frac{\sqrt{3}}{2}(z^2+2zx)-\frac{2\pi}{3}\right)\right) \, ,\\ \nonumber
&&\sum _{k=0}^{\infty}\frac{z^{3k+2}}{(3k+2)!}H_{3k+2}(x)=\, _{2}e^{-z^2+2zx}=\frac{1}{3}\left(  e^{-z^2+2zx}+2e^{\frac{z^2}{2}-zx}\cos \left(\frac{\sqrt{3}}{2}(z^2+2zx)+\frac{2\pi}{3}\right)\right) \,.
\end{eqnarray*}

\subsection{Mod n Hermite polynomials}
To describe wave functions for kaleidoscope states in coordinate representation for arbitrary $n$, we introduce mod $n$ Hermite polynomials. Generating function for these polynomials
\begin{eqnarray} \label{modngeneratingfunc1}
\sum _{s=0}^{\infty}\frac{z^{ns+k} }{(ns+k)!}H_{ns+k}(x)=\,\,_{k}e^{-z^2+2zx} \, ,
\end{eqnarray}
is defined by mod $n$ composite exponential functions;
\begin{eqnarray} \label{modngeneratingfunc2}
_{k}e^{-z^2+2zx}\equiv \frac{1}{n}\sum _{s=0}^{n-1} \bar{q}^{2ks} e^{-(q^{2s}z)^2+2(q^{2s}z)x}\,,
\end{eqnarray}
where in $(\ref{superpositionofnfunction})$, we have used $\displaystyle{f(z)=e^{-z^2+2zx}}$ , for $ 0\leq k \leq n-1\,.$

\section{Schr\"{o}dinger's mod 2 cat states}
The Schr\"{o}dinger cat states as an even and odd superposition of $| \alpha \rangle$ and $|-\alpha \rangle$ states, represent mod 2 kaleidoscope states. Here by mod 2 exponential functions, we calculate  explicitly several characteristics of these states as normalization constants, average number of photons, the uncertainty relations and coordinate representation. The normalization of Schr\"{o}dinger cat states $(\ref{notnormalizedcat1}),(\ref{notnormalizedcat2})$ is represented in terms of mod 2 exponential functions as,
\begin{eqnarray} \label{cat1}
|0\rangle_{{\alpha}}&=&\frac{_0 e^{\alpha\hat{a}^\dag} }{\sqrt{_0 e^{|\alpha|^2}}}|0\rangle = \frac{\cosh \alpha \hat{a}^\dag}{\sqrt{\cosh |\alpha|^2}} |0\rangle \quad(mod \,2) \, , \\ \label{cat2}
|1\rangle_{{\alpha}}&=&\frac{_1 e^{\alpha\hat{a}^\dag}}{\sqrt{_1 e^{|\alpha|^2}}}|0\rangle= \frac{\sinh \alpha \hat{a}^\dag}{\sqrt{\sinh |\alpha|^2}} |0\rangle\quad(mod \,2).
\end{eqnarray}
Average number of photons in these cat states is propotional to ratio of normalization constants,
\begin{eqnarray} \label{photonsofcat0}
{_{\alpha}}\langle0|\widehat{N}|0\rangle_{{\alpha}}&=&|\alpha|^2\frac{_1 e^{|\alpha|^2}}{_0 e^{|\alpha|^2}} = |\alpha|^2 \tanh |\alpha|^2\,
 ,  \\
 \label{photonsofcat1} {_{\alpha}}\langle1|\widehat{N}|1\rangle_{{\alpha}}&=&|\alpha|^2\frac{_0 e^{|\alpha|^2}}{_1 e^{|\alpha|^2}}= |\alpha|^2 \coth |\alpha|^2\,.
\end{eqnarray}
As easy to evaluate, asymptotically these numbers are approaching the usual coherent states number $|\alpha|^2:$
\begin{eqnarray*}
\lim_{|\alpha|\to\infty}{_{\alpha}}\langle0|\widehat{N}|0\rangle_{{\alpha}}= \lim_{|\alpha|\to\infty}{_{\alpha}}\langle1|\widehat{N}|1\rangle_{{\alpha}}
\approx |\alpha|^2=\langle\pm\alpha|\widehat{N}|\pm\alpha\rangle \, .
\end{eqnarray*}
In the limit $|\alpha|\rightarrow0$, we get number of photons in the so called Schr\"{o}dinger's kitten states:
\begin{eqnarray*}
\lim_{|\alpha|\to0}{_{\alpha}}\langle0|\widehat{N}|0\rangle_{{\alpha}}=0 ,\quad
\lim_{|\alpha|\to0}{_{\alpha}}\langle1|\widehat{N}|1\rangle_{{\alpha}}=1 .
\end{eqnarray*}
\\It is known that in contrast to coherent states, Schr\"{o}dinger's cat states are not satisfying minimum uncertainty relation, but instead
\begin{eqnarray*}
\left( \Delta \hat{q} \right) _{|0\rangle_{{\alpha}}} \left( \Delta \hat{p} \right) _{|0\rangle_{{\alpha}}}&=&\frac{\hbar}{2}
\sqrt{\left(1+2\,{_{\alpha}}\langle0|\widehat{N}|0\rangle_{{\alpha}}\right)-\left(\alpha^2+\overline{\alpha}^2\right)^2} \, , \\
\left( \Delta \hat{q} \right) _{|1\rangle_{{\alpha}}} \left( \Delta \hat{p} \right) _{|1\rangle_{{\alpha}}}&=&
\frac{\hbar}{2} \sqrt{\left(1+2\,{_{\alpha}}\langle1|\widehat{N}|1\rangle_{{\alpha}}\right)-\left(\alpha^2+\overline{\alpha}^2\right)^2} \, ,
\end{eqnarray*}
where average number of photons are given by $(\ref{photonsofcat0}),(\ref{photonsofcat1})$.
To write the  cat states in coordinate representation we use mod $2$ Hermite polynomials, which can be derived by following mod $2$ generating functions,
\begin{eqnarray} \label{coordinatecat1}
\langle x |0\rangle_{\alpha}=\frac{e^{-\frac{x^2}{2}}}{\pi^{1/4}\sqrt{\cosh|\alpha|^{2}}}\sum _{n=0}^{\infty}\frac{  H_{2k}(x)}{(2k)!} \left(\frac{\alpha}{\sqrt{2}}\right)^{2k} =\frac{e^{-\frac{x^2}{2}}}{\pi^{1/4}\sqrt{\,_{0}e^{|\alpha|^2}}}\,\,\,_{0}e^{-\frac{\alpha^2}{2}+\sqrt{2}\alpha x} \,,
\end{eqnarray}
\begin{eqnarray}\label{coordinateca2}
\langle x |1\rangle_{\alpha}=\frac{e^{-\frac{x^2}{2}}}{\pi^{1/4}\sqrt{\sinh|\alpha|^{2}}}\sum _{n=0}^{\infty}\frac{  H_{2k+1}(x)}{(2k+1)!} \left(\frac{\alpha}{\sqrt{2}}\right)^{2k+1}= \frac{e^{-\frac{x^2}{2}}}{\pi^{1/4}\sqrt{\,_{1}e^{|\alpha|^2}}}\,\,\,_{1}e^{-\frac{\alpha^2}{2}+\sqrt{2}\alpha x}\,.
\end{eqnarray}

\section{Trinity states}
As a first generalization of Schr\"{o}dinger cat states, we introduce the trinity states. If the cat states are associated with $q^4=1,$ the trinity states are related with $q^6=1\,,$ so that $\displaystyle{q^2=e^{i\frac{2\pi}{3}}}.$ This generalization is constructed by superposition of coherent states, rotated by angle $\frac{2\pi}{3}$ and associated with vertices of equilateral triangle. Then, the set of three orthonormal states $|0\rangle_{\alpha},|1\rangle_{\alpha}$ and $ |2\rangle_{\alpha}$ ,the trinity states, is
\begin{eqnarray*}
 |0\rangle_{\alpha}&=&e^{\frac{|\alpha|^{2}}{2}} \frac{| \alpha \rangle+|q\,^{2}\alpha\rangle+|q\,^{4}\alpha\rangle}
{\sqrt{3}\sqrt{e^{|\alpha|^{2}}+e^{{q}^{2}|\alpha|^{2}}+e^{{q}^{4}|\alpha|^{2}}}}
 =e^{\frac{|\alpha|^{2}}{2}} \frac{| \alpha \rangle + |q\,^{2}\alpha\rangle+|q\,^{4}\alpha\rangle }{3\sqrt{ _{0}e^{|\alpha|^2}{(mod \,3)}}} \, ,  \nonumber \\ \nonumber \\
 |1\rangle_{\alpha}&=& e^{\frac{|\alpha|^{2}}{2}} \frac{| \alpha \rangle+\overline{q}^{2}|q\,^{2}\alpha\rangle+\overline{q}^{4}|q\,^{4}\alpha\rangle}
{\sqrt{3}\sqrt{e^{|\alpha|^{2}}+\overline{q}^{2}e^{{q}^{2}|\alpha|^{2}}+\overline{q}^{4}e^{{q}^{4}|\alpha|^{2}}}}
= e^{\frac{|\alpha|^{2}}{2}} \frac{| \alpha \rangle +\overline{q}^{2} |q\,^{2}\alpha\rangle+
 \overline{q}^{4}|q\,^{4}\alpha\rangle }{3\sqrt{ _{1}e^{|\alpha|^2}(mod \,3)}}\, , \nonumber \\
|2\rangle_{\alpha}&=& e^{\frac{|\alpha|^{2}}{2}} \frac{| \alpha \rangle+\overline{q}^{4}|q\,^{2}\alpha\rangle+\overline{q}^{2}|q\,^{4}\alpha\rangle}
{\sqrt{3}\sqrt{e^{|\alpha|^{2}}+\overline{q}^{4}e^{{q}^{2}|\alpha|^{2}}+\overline{q}^{2}e^{{q}^{4}|\alpha|^{2}}}}
= e^{\frac{|\alpha|^{2}}{2}} \frac{| \alpha \rangle +\overline{q}^{4} |q\,^{2}\alpha\rangle+
 \overline{q}^{2}|q\,^{4}\alpha\rangle }{3\sqrt{ _{2}e^{|\alpha|^2}(mod \,3)}}\, .\nonumber
\end{eqnarray*}
By using definition of mod $3$ exponential functions, we can obtain trinity states in a compact form;
\begin{eqnarray*}
|0\rangle_{{\alpha}}=\frac{_0 e^{\alpha\hat{a}^\dag} }{\sqrt{_0 e^{|\alpha|^2}}}|0\rangle\, , \quad
|1\rangle_{{\alpha}}=\frac{_1 e^{\alpha\hat{a}^\dag}}{\sqrt{_1 e^{|\alpha|^2}}}|0\rangle\, , \quad
|2\rangle_{{\alpha}}=\frac{_2 e^{\alpha\hat{a}^\dag}}{\sqrt{_2 e^{|\alpha|^2}}}|0\rangle \qquad (mod\,\,3).
\end{eqnarray*}

\subsection{Number of photons and  uncertainty relations}
For calculating number of photons in trinity states, it is convenient to apply annihilation operator $\hat{a}$ to the states $|0\rangle_{{\alpha}},|1\rangle_{{\alpha}}$ and $|2\rangle_{{\alpha}}$. The operator $\hat{a}$ acts on these states as cyclic permutation and
number of photons is determined by ratio of two consecutive mod $3$ exponential functions,
\begin{eqnarray} \label{photonnumberfortrinity0}
{_{\alpha}}\langle0|\widehat{N}|0\rangle_{{\alpha}}&=&|\alpha|^2\left(\frac{_2 e^{|\alpha|^2}}{_0 e^{|\alpha|^2}}\right)
 \,, \\ \label{photonnumberfortrinity1}
{_{\alpha}}\langle1|\widehat{N}|1\rangle_{{\alpha}}&=&|\alpha|^2\left(\frac{_0 e^{|\alpha|^2}}{_1 e^{|\alpha|^2}}\right)
 \, ,\\ \label{photonnumberfortrinity2}
{_{\alpha}}\langle2|\widehat{N}|2\rangle_{{\alpha}}&=&|\alpha|^2\left(\frac{_1 e^{|\alpha|^2}}{_2e^{|\alpha|^2}}\right)\,.
\end{eqnarray}
For small number of photons, we get in  the limit $|\alpha|^2\,\rightarrow\,0 $;
\begin{eqnarray*}
\lim_{|\alpha|^2\to0}{_{\alpha}}\langle0|\widehat{N}|0\rangle_{{\alpha}}=0 \, , \quad
\lim_{|\alpha|^2\to0}{_{\alpha}}\langle1|\widehat{N}|1\rangle_{{\alpha}}=1\, , \quad
\lim_{|\alpha|^2\to0}{_{\alpha}}\langle2|\widehat{N}|2\rangle_{{\alpha}}=2 \, .
\end{eqnarray*}
 The uncertainty relations can be expressed explicitly  in terms of these numbers,
\begin{eqnarray*}
\left( \Delta \hat{q} \right) _{|k\rangle_{{\alpha}}} \left( \Delta \hat{p} \right) _{|k\rangle_{{\alpha}}}&=&\frac{\hbar}{2}
\left(1+2\,{_{\alpha}}\langle k|\widehat{N}|k\rangle_{{\alpha}}\right) \, ,
\end{eqnarray*}
where ${_{\alpha}}\langle k|\widehat{N}|k\rangle_{{\alpha}} \, , 0\leq k \leq2$ are given by $(\ref{photonnumberfortrinity0})$-$(\ref{photonnumberfortrinity2})$. In the limiting case $|\alpha|^2\,\rightarrow\,0 $, uncertainty is growing with states number $k.$

\section{Kaleidoscope of quantum coherent states}
As a generalization of previous results, here we consider superposition of $n$ coherent states, which are belonging to vertices of regular $n$-polygon and are rotated by angle $\frac{2\pi}{n},$ related with primitive roots of unity $q^{2n}=1.$
This kaleidoscope of quantum coherent states can be described by simple formula using mod $n$ exponential,
\begin{eqnarray*}
|k \rangle_{\alpha}=\frac{_{k}e^{ \alpha \hat{a}^{\dagger}}}
{\sqrt{_{k}e^{|\alpha|^2}}}| 0 \rangle \quad (mod \, n), \,\,\,0\leq k \leq n-1. \label{modnkaleidoscope}
\end{eqnarray*}

\subsection{Number of photons in kaleidoscope states}
It allows us to calculate number of photons by simple application of annihilation operator. The kaleidoscope of quantum states can be  generated by annihilation operator $\hat{a}$ acting as cyclic permutation of these states;
\begin{eqnarray}
\hat{a}|0\rangle_{{\alpha}}=\alpha \, \sqrt{\frac{_{n-1}e^{|\alpha|^2}}{_{0}e^{|\alpha|^2}}}|n-1\rangle_{{\alpha}}  \label{annihilationtozeroalpha}\, ,\quad \quad\hat{a}|k\rangle_{{\alpha}}=\alpha \,  \sqrt{\frac{_{k-1}e^{|\alpha|^2}}{_{k}e^{|\alpha|^2}}} |k-1\rangle_{{\alpha}}\, . \label{annihilationtokalpha}
\end{eqnarray}
By taking norm of these states, we get average number of photons as,$ 1\leq k \leq n-1,$
\begin{eqnarray*}
{_{\alpha}}\langle 0|\widehat{N}|0 \rangle_{{\alpha}}=|\alpha|^2\left(\frac{_{n-1}e^{|\alpha|^2}}{_{0}e^{|\alpha|^2}}\right)\, ,\quad
{_{\alpha}}\langle k|\widehat{N}|k \rangle_{{\alpha}}=|\alpha|^2\left(\frac{_{k-1}e^{|\alpha|^2}}{_{k}e^{|\alpha|^2}}\right).
\end{eqnarray*}
Asymptotically, for small occupation numbers they approach the integer values
\begin{eqnarray*} \label{numberofphotonslimitinkaleidoscopeasalphagoeszero}
\lim_{|\alpha|\to0}{_{\alpha}}\langle k|\widehat{N}|k\rangle_{{\alpha}}=k \, , 0\leq k \leq n-1\, .
\end{eqnarray*}

\subsection{Heinsenberg uncertainty relation for kaleidoscope states}
 The following uncertainty relations are valid for $n\geq3;$
\begin{eqnarray*} \label{uncertaintyforrelationforkaleidosopestates}
\left( \Delta \hat{q} \right) _{|k\rangle_{{\alpha}}} \left( \Delta \hat{p} \right) _{|k\rangle_{{\alpha}}}= \frac{\hbar}{2} \left(1+2|\alpha|^2 \,\frac{_{k-1}e^{|\alpha|^2}}{_{k}e^{|\alpha|^2}}\right) \, ,
\end{eqnarray*}
where
\begin{eqnarray*}
\left( \Delta \hat{q} \right) _{|k\rangle_{{\alpha}}}\equiv \left( \Delta \hat{p} \right) _{|k\rangle_{{\alpha}}}=\sqrt{ \frac{\hbar}{2} \left(1+2|\alpha|^2 \,\frac{_{k-1}e^{|\alpha|^2}}{_{k}e^{|\alpha|^2}}\right)} \,.
\end{eqnarray*}
It is noticed that the form of variance here is different from the cat states $(n=2)$, since the cat states are eigenstates of operator $\hat{a}^{2}.$
The uncertainty relation for kaleidoscope states $|k\rangle_{{\alpha}}$  have the following limit,
\begin{eqnarray*}
\lim_{|\alpha|^2\to0} \left( \Delta \hat{q} \right) _{|k\rangle_{{\alpha}}} \left( \Delta \hat{p} \right) _{|k\rangle_{{\alpha}}} = \frac{\hbar}{2}(2k+1) \quad , 0\leq k \leq n-1\, .
\end{eqnarray*}
The above relation coincides with spectrum of harmonic oscillator with finite number of energy levels
\begin{eqnarray*}
E_k=\hbar\left(k+\frac{1}{2}\right) \quad ,\, 0\leq k \leq n-1\,.
\end{eqnarray*}

\subsection{Coordinate representation of kaleidoscope states}
The wave function for kaleidoscope of quantum coherent states $|k \rangle_{{\alpha}}\, , 0\leq k \leq n-1\,,$ in coordinate representation is given by
\begin{eqnarray*}
\langle x |k\rangle_{\alpha}&=&\frac{e^{-\frac{x^2}{2}}}{\pi^{1/4}\sqrt{\,_{k}e^{|\alpha|^2}}}\sum _{s=0}^{\infty}\frac{H_{ns+k}(x) }{(ns+k)!}\left(\frac{\alpha}{\sqrt{2}}\right)^{ns+k} \,,
\end{eqnarray*}
and with $(\ref{modngeneratingfunc1}),(\ref{modngeneratingfunc2})$, it appears as superposition of Gaussian wave functions
\begin{eqnarray*}
\langle x |k\rangle_{\alpha}&=&\frac{e^{-\frac{x^2}{2}}}{\pi^{1/4}\sqrt{\,_{k}e^{|\alpha|^2}}}\,\,\,_{k}e^{-\frac{\alpha^2}{2}+\sqrt{2}\alpha x} \,,
\end{eqnarray*}
where mod n composite exponential functions are defined by $(\ref{superpositionofnfunction})$ with $\displaystyle{f(\alpha)=\,_{k}e^{-\frac{\alpha^2}{2}+\sqrt{2}\alpha x}}$ ,
\begin{eqnarray*}
_{k}e^{-\frac{\alpha^2}{2}+\sqrt{2}\alpha x}\equiv \frac{1}{n}\sum _{s=0}^{n-1} \bar{q}^{2ks} e^{-\frac{1}{2}(q^{2s}\alpha)^2+\sqrt{2}(q^{2s}\alpha)x}
\end{eqnarray*}
In particular case of mod 4 quartet states with $q^8=1$, the wave functions are calculated as:
\begin{eqnarray*}
&&\langle x |0\rangle_{\alpha}= \frac{e^{-\frac{x^2}{2}}}{\sqrt{2}\pi^{1/4}} \frac{e^{-\frac{\alpha^2}{2}}\cosh (\sqrt{2}\alpha x)+e^{\frac{\alpha^2}{2}}\cos (\sqrt{2}\alpha x)}{\sqrt{\cosh |\alpha|^2+\cos |\alpha|^2 }} \,, \\
&&\langle x |1\rangle_{\alpha}= \frac{e^{-\frac{x^2}{2}}}{\sqrt{2}\pi^{1/4}}  \frac{e^{-\frac{\alpha^2}{2}}\sinh (\sqrt{2}\alpha x)+e^{\frac{\alpha^2}{2}}\sin (\sqrt{2}\alpha x) }{\sqrt{\sinh |\alpha|^2+\sin |\alpha|^2}},\\
&&\langle x |2\rangle_{\alpha}=\frac{e^{-\frac{x^2}{2}}}{\sqrt{2}\pi^{1/4}} \frac{e^{-\frac{\alpha^2}{2}}\cosh (\sqrt{2}\alpha x)-e^{\frac{\alpha^2}{2}}\cos (\sqrt{2}\alpha x)}{\sqrt{\cosh |\alpha|^2-\cos |\alpha|^2}}\,, \\
&&\langle x |3\rangle_{\alpha}= \frac{e^{-\frac{x^2}{2}}}{\sqrt{2}\pi^{1/4}}  \frac{e^{-\frac{\alpha^2}{2}}\sinh (\sqrt{2}\alpha x)-e^{\frac{\alpha^2}{2}}\sin (\sqrt{2}\alpha x)}{\sqrt{\sinh |\alpha|^2-\sin|\alpha|^2 }}\,.
\end{eqnarray*}
Corresponding probabilities with mod 4 symmetries for $\alpha=1+i$ are shown in Figures 1-4.
\clearpage
\begin{figure}[h]
\hspace{3.4pc}\begin{minipage}{14pc}
\includegraphics[scale=0.55]{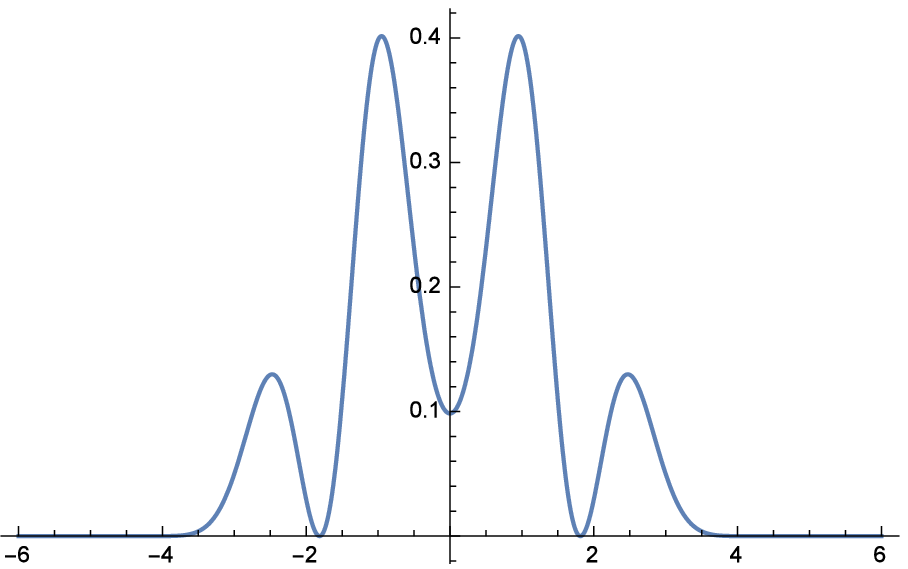}
\caption{Probability for mod 4 $|0\rangle$ state}
\end{minipage}\hspace{4pc}%
\begin{minipage}{14pc}
\includegraphics[scale=0.55]{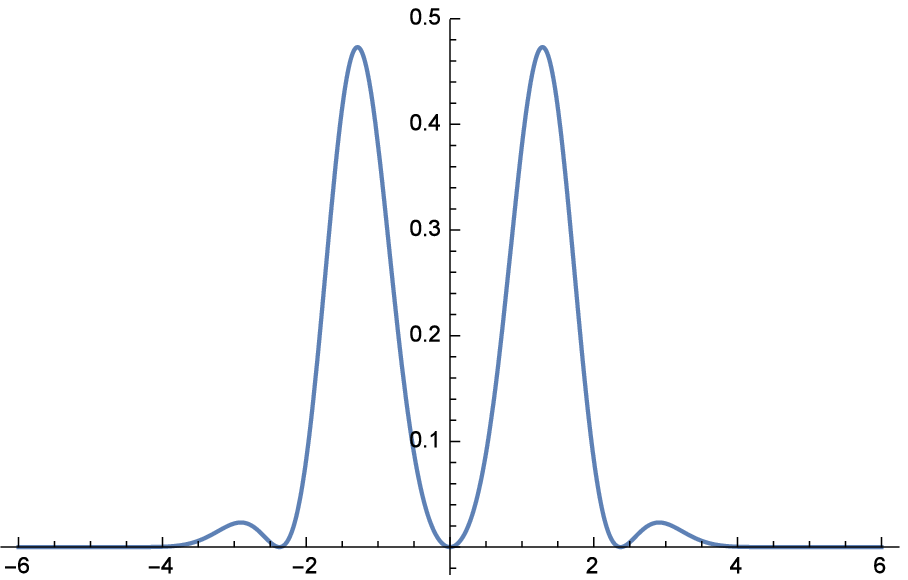}
\caption{Probability for mod 4 $|1\rangle$ state}
\end{minipage}
\end{figure}
\begin{figure}[h]
\hspace{3.4pc}\begin{minipage}{14pc}
\includegraphics[scale=0.55]{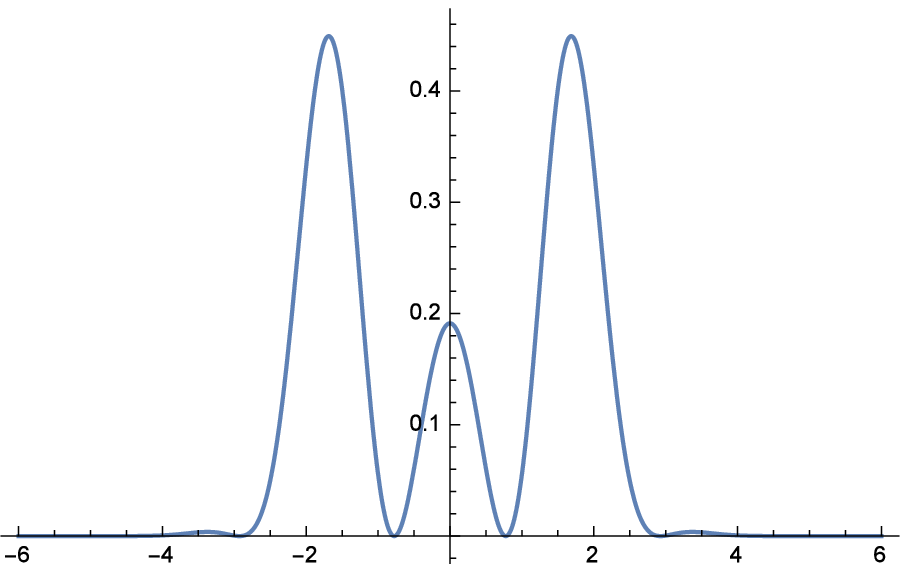}
\caption{Probability for mod 4 $|2\rangle$ state}
\end{minipage}\hspace{4pc}%
\begin{minipage}{14pc}
\includegraphics[scale=0.55]{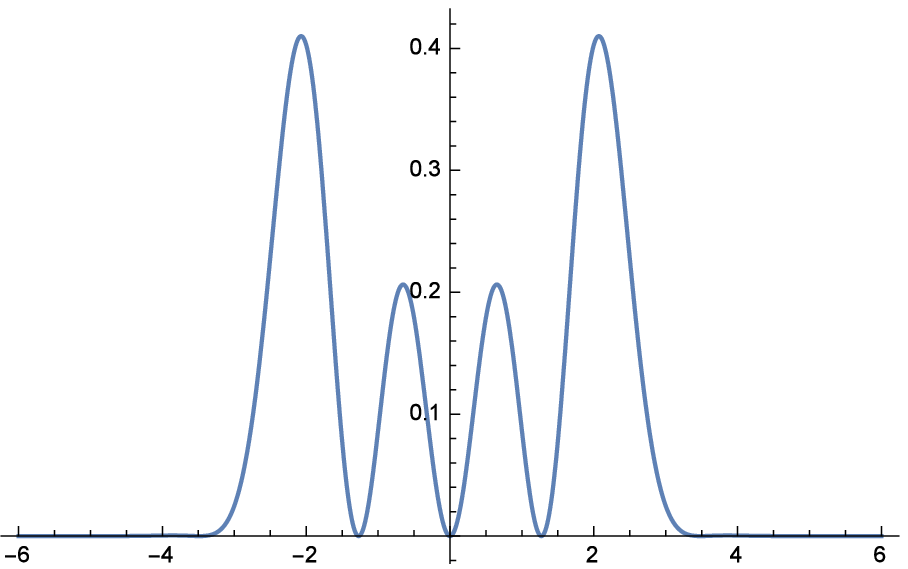}
\caption{Probability for mod 4 $|3\rangle$ state}
\end{minipage}
\end{figure}

As we can see, for even states $|0\rangle$ and $|2\rangle$ probabilites at origin are not zero, while for odd states $|1\rangle$ and $|3\rangle$, these probabilities vanish at origin. More detail analysis of probability distribution for different mod $n$ kaleidoscope states is under investigation.

\section{Acknowledgements} This work is supported by TUBITAK grant 116F206.



\begin{thebibliography}{99}
\bibitem{kocakpashaev} Pashaev O K and Kocak A 2018 \textit{Symmetries, Differential Equations and Applications\,,Springer} vol 266, ed V G Kac and
P J Olver \textit{et al} (Zurich, Springer) pp 179-199
\bibitem{bb} Bialynicka-Birula Z 1968 \textit{Phy Rev} \textbf{173} 1207
\bibitem{Stoler} Stoler D 1971 \textit{Phy Rev D} \textbf{4} 2309
\bibitem{Spiridonov} Spiridonov V V 1995 \textit{Phy Rev A} \textbf{52} 1909
\bibitem{ungar} Ungar A 1982 \textit{The American Mathematical Monthly} \textbf{89} 688-691
\end{thebibliography}
\end{document}